\documentclass{aa}
\usepackage{amsmath,amssymb}
\usepackage{mathtools}
\usepackage{graphicx}
\usepackage{subfigure}
\usepackage[colorlinks=true,citecolor=blue,linkcolor=blue,urlcolor=blue,breaklinks]{hyperref}
\usepackage{txfonts}
\usepackage{CJK}
\usepackage{multirow,diagbox,ulem,comment}
\usepackage{orcidlink}

\newcommand{\psr}{PSR\,J1906+0746\xspace}
\newcommand{\msun}{\ifmmode\mbox{M}_{\odot}\else$\mbox{M}_{\odot}$\fi\xspace}

\begin{document}
\begin{CJK*}{UTF8}{gbsn}

\title{A deep search for radio pulsations from the 1.3\,{\msun} compact-object binary companion of young pulsar \psr}
\titlerunning{A pulsar search for the companion of \psr}

\def\orcid#1{\unskip$\orcidlink{#1}$}

\author{
Yuyang~Wang (王宇阳) \inst{1}\fnmsep\thanks{\tt y.wang3@uva.nl, leeuwen@astron.nl} \orcid{0000-0002-3822-0389}\
 \and
Joeri~van Leeuwen \inst{2}\fnmsep$^\star$ \orcid{0000-0001-8503-6958}\ 
}
\authorrunning{Yuyang Wang and Joeri~van Leeuwen}
\institute{Anton Pannekoek Institute for Astronomy, University of Amsterdam, Science Park 904, 1098 XH Amsterdam, The Netherlands 
\and
ASTRON, the Netherlands Institute for Radio Astronomy, Oude Hoogeveensedijk 4, 7991 PD, Dwingeloo, The Netherlands
}


\abstract{Double pulsar systems offer unrivaled advantages for the study of both astrophysics and fundamental physics.
But only one has been visible: PSR~J0737$-$3039; and its component pulsar B has now rotated out of sight due to the general-relativistic effect of geodetic precession.
We know, though, that these precession cycles can also pivot pulsars into sight, and that this precession occurs at similar strength in {\psr}.
That source is a young, unrecycled radio pulsar, orbiting a compact object with mass $\sim $1.32\,\msun.
This work presents a renewed campaign to detect radio pulsations from this companion, two decades after the previous search. 
Two key reasons driving this reattempt are the possibility that the companion radio beam has since precessed into our line of sight, and the improved sensitivity now offered by the FAST radio telescope. 
In 28 deep observations, we did not detect a credible companion pulsar signal.
After comparing the possible scenarios, we conclude the companion is still most likely a pulsar that is not
pointing at us. 
We next present estimates for the sky covered by such systems throughout their precession cycle. 
We find that for most system geometries, the all-time beaming fraction is unity, i.e., observers in any direction can see the system at some point. 
We conclude it is still likely that \psr will be visible as a double pulsar in the future. }

\maketitle

\end{CJK*}

\section{Introduction} 
Double pulsar systems are highly valuable for strong-field tests of general relativity (GR) and for stellar evolution studies \citep[as reviewed in, e.g.,][]{Kramer2008}. 
To date, the only double pulsar system that is (or more accurately: has been) visible is PSR J0737$-$3039A/B.
It consists of a recycled millisecond pulsar (MSP), PSR J0737$-$3039A \citep{Burgay2003}; and a young, normal pulsar, PSR J0737$-$3039B \citep{Lyne2004}. 
Observations of PSR J0737$-$3039A/B helped improve our understanding of the pulsar wind, of the plasma surrounding it, of the orbital modulation, and of the pulsar magnetosphere \citep{Lyne2004}. 
It further enabled tests of GR and of modified gravity, by precisely measuring the post-Keplerian (PK) parameters \citep{Kramer2006}. The spin precession rates of both A and B were constrained \citep{Perera2010}. 
It furthered our understanding of the origin and evolution of double neutron star (DNS) systems, and improved our estimates of the DNS coalescence rate for ground-based gravitational-wave (GW) detectors.

The huge scientific potential of such double pulsar systems prompts us to look for more, even if (or, especially because) they are rare. 
One important factor in the visibility of double pulsar systems, is geodetic precession \citep{dr74}.
This GR effect causes changes in the direction in which the pulsar beams are emitted, on the timescales of only years.
That causes great variation in the visibility and detectability of these systems, from one decade to the next.
PSR J0737$-$3039B, for example, is no longer visible \citep{Noutsos2020}. 
On the other hand, systems that were single-pulsar at detection may since have temporarily changed to double-pulsar systems. 
Out of a number of such candidate binaries, one very promising system is \psr \citep{Lorimer2006,lcl+06,vanLeeuwen2015}. 
\psr, a young pulsar of age $\sim$ 112 \,kyr, is in a very short 3.98-hr orbit around another compact object. Since the discovery in the PALFA survey with Arecibo in 2004, precise timing solutions \citep{kasi11,vanLeeuwen2015,Vleeschower2024} have determined the exact parameters of the masses of the system, and established the companion mass to be 1.32\,\msun.
As noted initially by \citet{Lorimer2006} and mentioned again later by \citet{YangYY2017}, the \psr system shares similarities with PSR 0737$-$3039A/B across many aspects.
These include important measures such as the neutron star masses, eccentricity, orbital period, geodetic precession rate, and the strong magnetic field of the young pulsar in both systems.
Intriguingly, the recycled pulsars in such systems are a priori more likely to be seen, as they generally have higher radio flux densities and longer lifetimes than normal pulsars \citep{Lyne2012,YangYY2017}. And indeed PSR J0737$-$3039A is visible. But no recycled pulsar has yet been observed in the \psr system.

Earlier deep searches with Arecibo found no radio pulsations from the companion to \psr. This indicates that the companion could be either a white dwarf (WD), a pulsar whose radio beam does not intersect our line-of-sight (LOS), or a faint pulsar with a radio luminosity below 0.1\,mJy\,kpc$^2$ \citep{Lorimer2006}.
A targeted search with Chandra on \psr did not detect any significant signal from the system either (i.e., from either the young pulsar, the MSP, or their interaction; \citealt{Kargaltsev2009}).
Nevertheless, the high geodetic precession rate of $\sim$ 2.19$^\circ$\,yr$^{-1}$ (predicted by GR; see the equations in \citealt{Kramer2009} and \citealt{WangThesis2025}) makes it possible that the radio beam of the presumptive companion pulsar will precess into our LOS in the future (unless the companion spin axis is aligned with the total orbital angular momentum axis of the system, as is the case for PSR J0737$-$3039A; \citealt{Ferdman2013}).
Therefore it is meaningful to now re-examine the companion visibility, 18 years after the previous searches. 
For this, we use $>$2 years of highly sensitive FAST observations from our \psr campaign (Wang et al.~2025, in prep.), taken between March 2022 and July 2024.

This paper is organized as follows. In Sect.~2, we will introduce the FAST observations of \psr. The data reduction, companion pulsar searching pipeline based on \texttt{PRESTO} and orbital demodulation methods and analysis of the possible companion signal will be presented in Sect.~3. A discussion on the implications is given in Sect.~4.

\section{Observations and data reduction}
\subsection{Observations}
The Five-hundred-meter Aperture Spherical radio Telescope (FAST; \citealt{Jiang2020}), located in Guizhou, China, is the largest and most sensitive single dish radio telescope in the world. 
We analyzed 28 observations taken between March 2022 and July 2024 (Wang et al.~2025, in prep.).
With the 19-beam (central) receiver, we used the Tracking mode to observe \psr and the OnOff mode to observe the flux calibrator PKS2209+080. The bandwidth of the FAST pulsar backend is 500\,MHz around a central frequency 1250\,MHz. The number of channels is 4096. The data is recorded in the search mode with a sampling time of 49.152\,$\mu$s. 
During the initial 1 minute of data taking on the pulsar the noise diode continued to fire, thus tying the observation to the flux calibration source. 
Offline, the data are cleaned from radio frequency interference (RFI), incoherently dedispersed, and polarization and flux calibrated.

\subsection{Data reduction}
\label{sec:datareduc}
We used \texttt{PRESTO}\footnote{\url{https://github.com/scottransom/presto}} \citep{Ransom2011} to produce the time
series from the filterbank files (i.e., dedisperse and integrate over all frequencies), and to next convert the time series to power spectra using fast Fourier transforms (FFTs), and search these.

We started by employing \texttt{rfifind} to find and remove the most prominent RFI.
We next identified topocentric, periodic RFI without dispersion measure (DM).
The frequencies of this RFI and of \psr are added to so-called zap list of frequencies, such that those can be blanked at subsequent stages.
We then used \texttt{prepsubband} to dedisperse the time series using DM=217.7508 pc\,cm$^{-3}$ and correct it such that it is valid for the Solar system barycenter.
These barycentered time series are input to the orbital demodulation, described in the next subsection. 
The outputs are time series that are corrected for the frame of motion of the companion. 
These we FFTed and corrected for excess red noise.
Finally we searched the resulting power spectra for periodic candidates\footnote{We defer a single-pulse search to potential future work. While a number of MSPs emit giant pulses (GPs; see, e.g., \citealt{Bilous2015} and references therein), all were discovered through their periodicity. One benefit of single pulse searches is that they do not require orbital demodulation; but the downside is that GPs are often narrow and their S/N would be much reduced in our 49.152\,$\mu$s samples. }
using \texttt{accelsearch}, summing \mbox{1--16} harmonics to cover both narrow and wide pulses. As the orbits had already been demodulated, we did not employ further acceleration searching in this step.
The highest signal-to-noise candidates are analyzed further as described in Sect.~\ref{sec:view}.

\subsection{Orbital demodulation}
The system is a tight binary, and the companion is continuously and strongly accelerated.
Our $\sim$2-hr observations span a significant part of the 3.98-hr orbit.
This ever-changing Doppler shift of the companion spin frequency throughout the orbit means any stable companion periodicity will be washed out in the observed data. 
But for systems with well-measured parameters, this smearing can be counteracted by taking out the expected orbital motion from the barycentered, dedispersed time series \citep{Lyne2004}.
In this study we use \texttt{pysolator}\footnote{\url{https://github.com/alex88ridolfi/PYSOLATOR}} \citep{Ridolfi2020} for this purpose. 
It removes the R{\o}mer delay predicted for the companion orbit, and resamples the time series, to transform the signal to the center of mass of the binary system. 

Essential input to this transformation is the exact ephemeris of the known pulsar, including the eccentricity of 0.085, and the mass fraction $q$.
The latter needed to calculate the semi-major axis $x_\mathrm{c}$ for the companion, from that of the pulsar $x_\mathrm{p}$, using $x_\mathrm{c} = q x_\mathrm{p}$.
For \psr, the pulsar mass $m_\mathrm{p} = 1.291\pm 0.006\,\msun$ and the companion mass $m_\mathrm{c} = 1.322\pm 0.006$\,\msun \citep{vanLeeuwen2015}.
Therefore, the $q$ range covering this 1$\sigma$ mass uncertainty must be [0.952, 1.002]. 
To remain completely sensitive to companion spin frequencies as high as 1000\,Hz in even our longest, 4-hr observations, the step in $q$ must be at most 0.0001.
Using the ephemeris from the long-term timing of \psr by Vleeschower et al.~(2025, in prep.) and the $q$ range described above, we produced demodulated companion trial time series for each observation.

\section{Results}
\subsection{Candidate vetting}
\label{sec:view}

\begin{figure}
    \includegraphics[width=\linewidth]{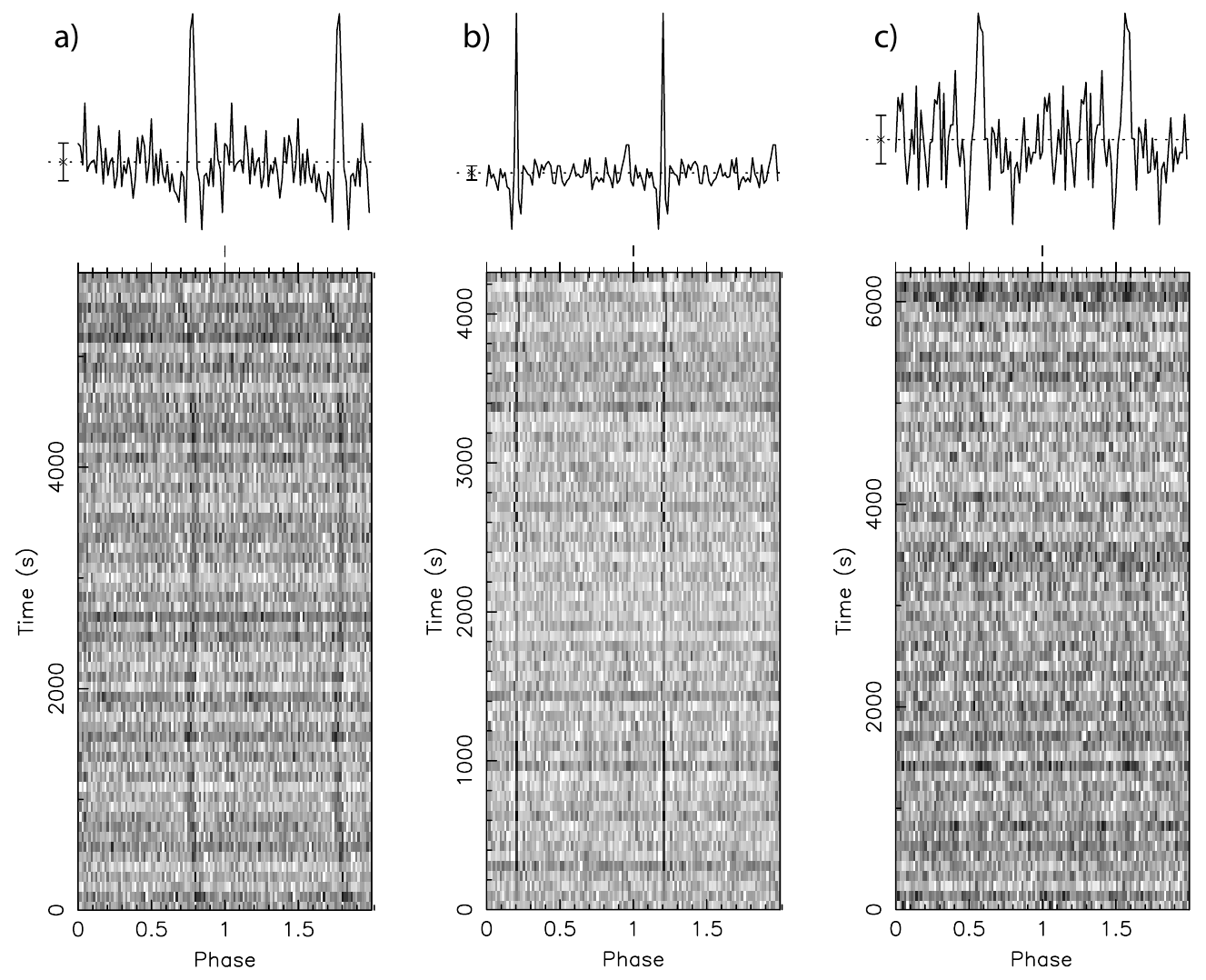}
    \caption{In a \texttt{prepfold} diagnostic plot,
    a candidate stands out when it displays pulsar-like integrated profile (top subpanels; two rotations shown) and a near-constant period once demodulated at the trial companion orbit (bottom subpanels)
    \textit{Left (a)}: the 2.059\,s-period candidate. \textit{Middle (b)}: the 4.019\,s candidate. \textit{Right (c)}: the 42\,ms candidate. }
    \label{Fig: prepfold}
\end{figure}

For each of our 28 observations, we produced characteristic (``\texttt{prepfold}'') plots for the candidates with the highest period signal-to-noise ratio (S/N).
A total of $4 \times 10^4$ such candidates were visually inspected. For each good candidate, we folded the original raw data, now without first integrating over frequency, to examine the frequency domain structure.
Pulsars are generally broadband while RFI is usually narrowband.

\subsection{Candidates}
\label{sec:cands}
\subsubsection{The 2.059\,s candidate}
\begin{figure}
    \includegraphics[width=\linewidth]{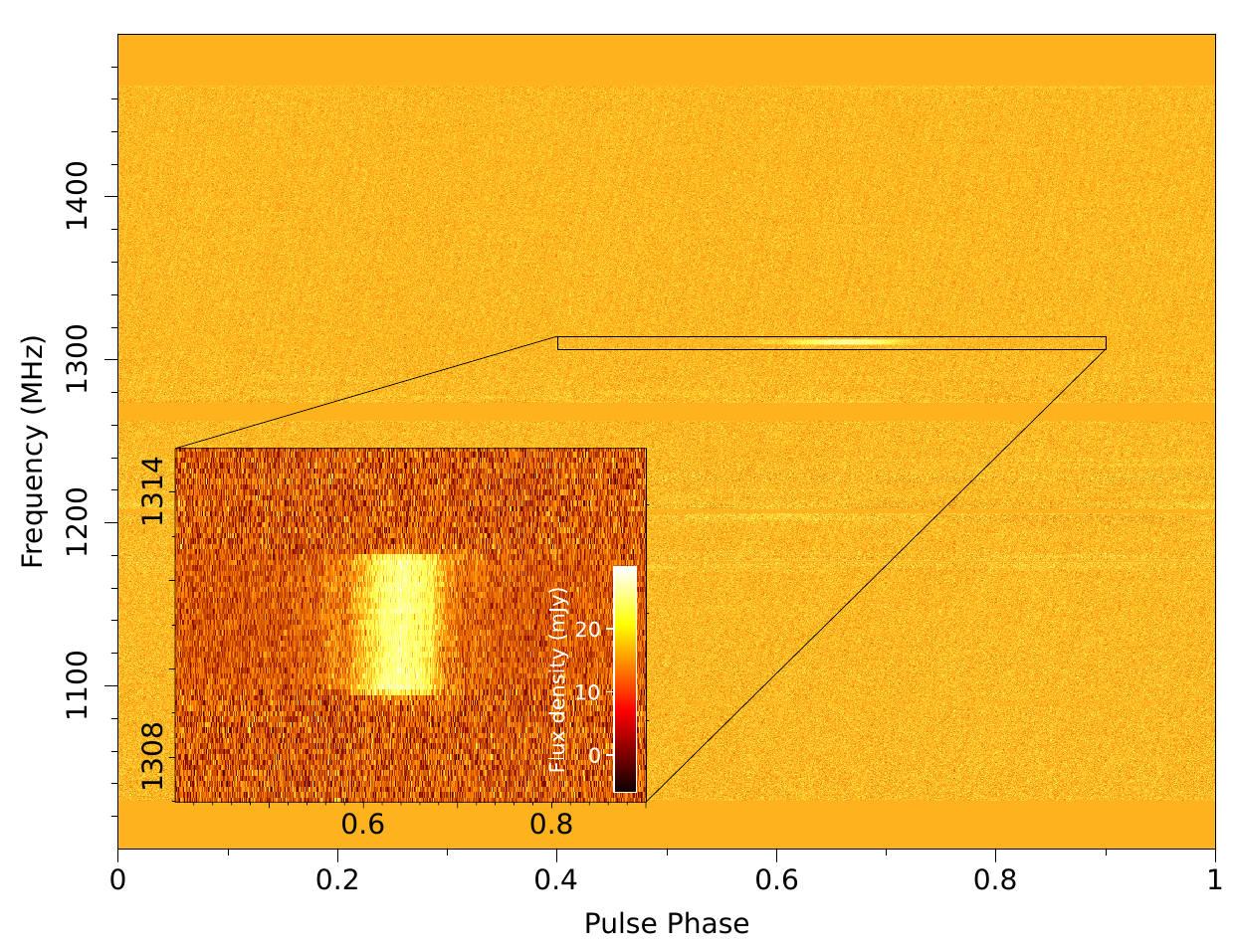}
    \caption{The 2.059\,s candidate flux density against frequency and pulse phase, folded with \texttt{dspsr}.
    The zoom-in plot shows the origin of the falsified candidate: it is narrowband RFI around 1310\,MHz.}
    \label{Fig: Cand-wZoom-2059ms}
\end{figure}

We found a strong 2.0594338\,s-period candidate, first in the 20230716 observation (see Fig.~\ref{Fig: prepfold}a), and soon thereafter confirmed at 4 other epochs.
The period does not agree with the expected MSP though, and would require a non-standard evolution scenario if proven to be real.
After folding the candidate from the raw data, we found it to be a narrow-band periodic signal in the 1309.4$-$1312.6\,MHz frequency band (Fig.~\ref{Fig: Cand-wZoom-2059ms}). We conclude this is RFI, likely arising from a radar system or a navigation satellite.
Another two candidates at other periods were subsequently found to stem from the same frequency band. 
This long-duration, narrow-band RFI, is too weak to be clipped by \texttt{rfifind} and \texttt{accelsearch}, which means it will likely appear in other FAST pulsar search projects.
Such studies are advised to incorporate \texttt{RFIClean} \citep{Maan2021} in their search pipeline. 

\subsubsection{The 4.018$-$4.045\,s candidate}
\begin{figure}
    \includegraphics[width=\linewidth]{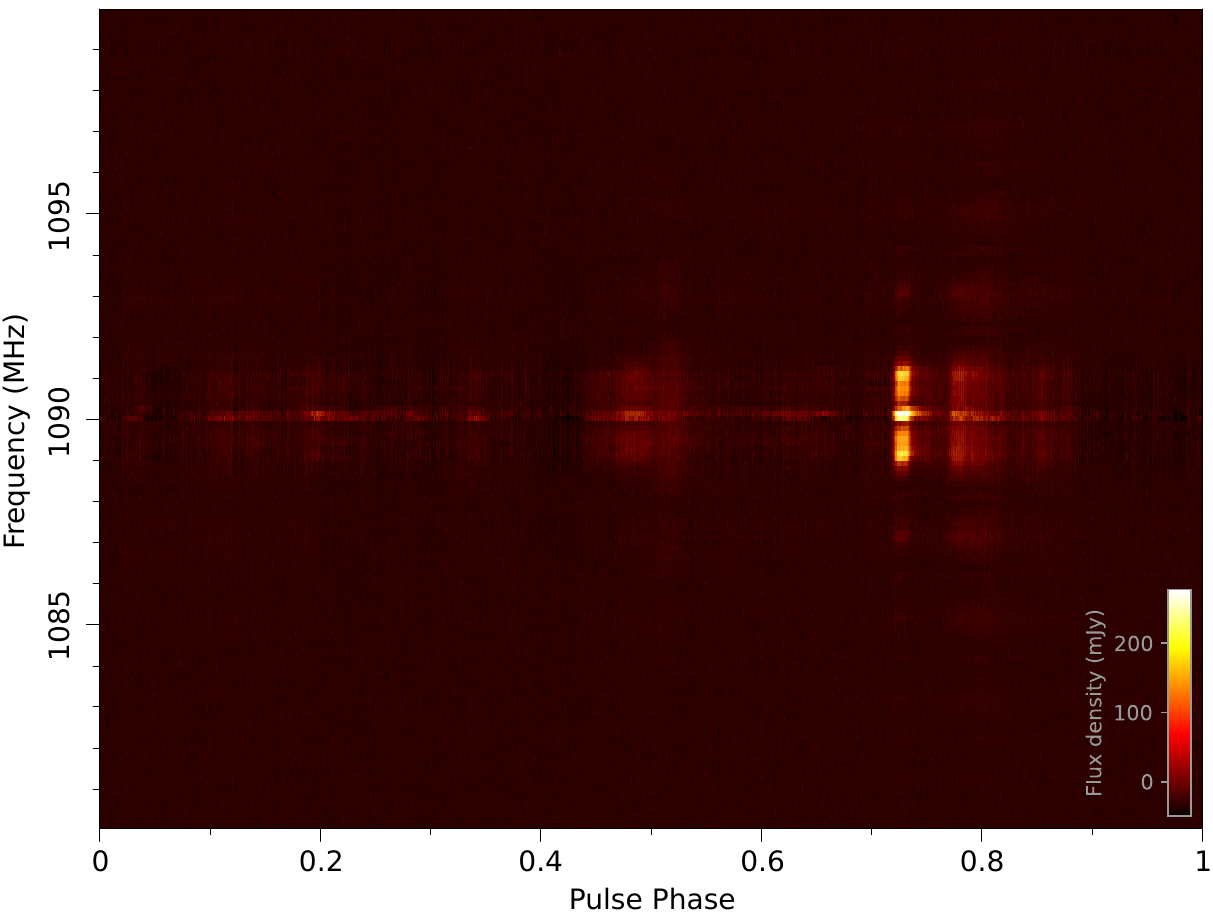}
    \caption{A zoom-in on the frequency range around the 4.019\,s candidate.
    Shown is flux density against frequency and pulse phase. }
    \label{Fig: Cand-wZoom-4018ms}
\end{figure}

A bright, apparently modulated candidate with a spin period of 4.033\,s was initially found in the 20231016 observation. At first, it was considered to be the 28th harmonic of {\psr}, but further investigation found candidates with similar spin periods in a number of other observations that were clearly not the harmonic.
At its highest, the candidate reaches a S/N of 24\,$\sigma$ (see Fig.~\ref{Fig: prepfold}b). 
It shares the same strong and weak points as the previous candidate. 
And again in the raw data it turns out to be periodic RFI, now around 1090\,MHz, with complex frequency-phase structure (Fig.~\ref{Fig: Cand-wZoom-4018ms}); based on the frequency these are possibly aircraft transponders.

\subsubsection{The 42\,ms candidate}
\label{sec:cands:42}
\begin{figure}
    \includegraphics[width=\linewidth]{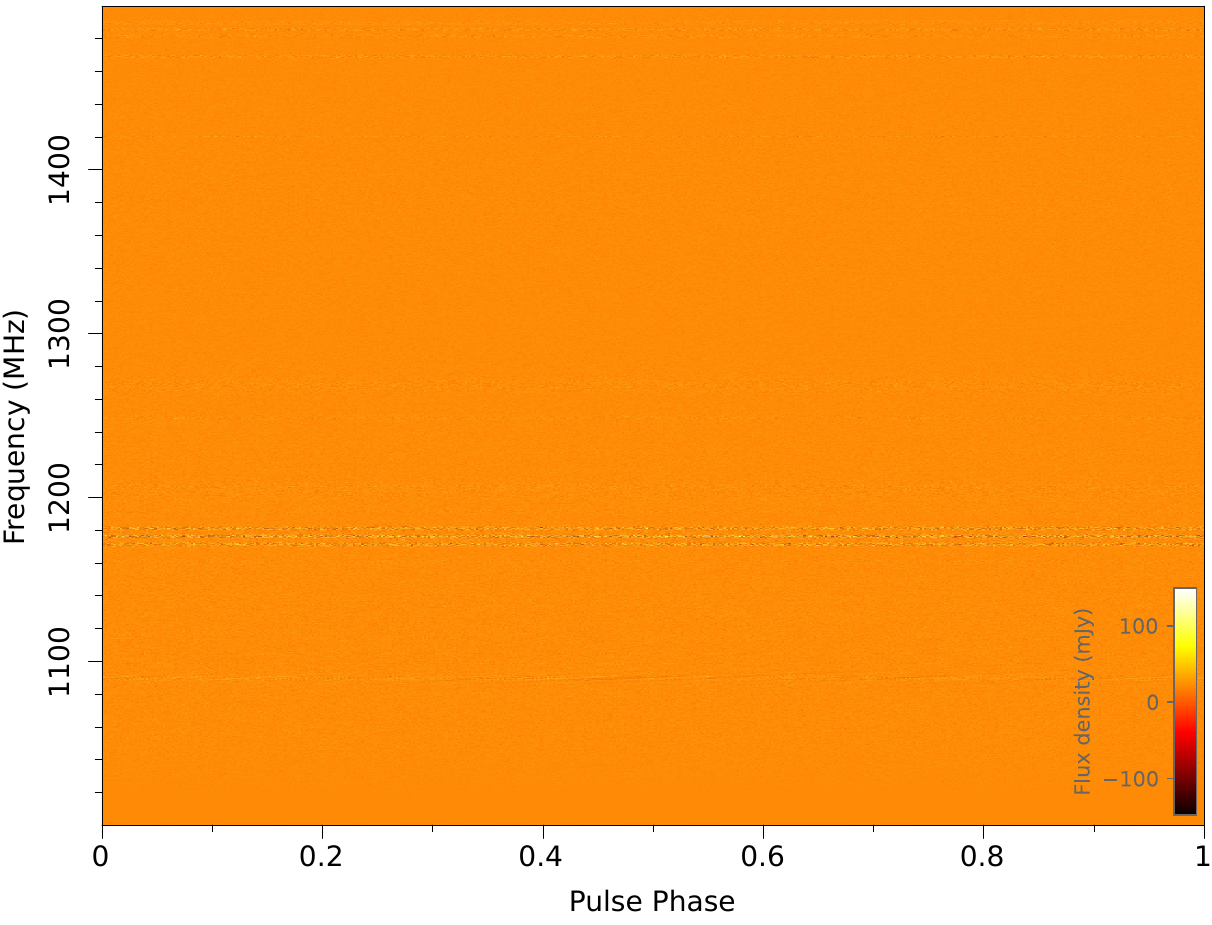}
    \caption{Frequency-phase plot of the flux density for the 42\,ms candidate, folded taking into account the effects of the implied orbit. No significant periodic signal is seen.}
    \label{Fig: Cand-42ms}
\end{figure}

A faster candidate with period 41.9379\,ms was detected in the dedispersed time series of observation 20221110 (see Fig.~\ref{Fig: prepfold}c) at 8$\sigma$. 
This period fits well with the mildly recycled pulsar that is expected to accompany {\psr} (see \citealt{vanLeeuwen2015} for a list of such systems). 
The candidate is especially convincing because its period (denoted by the persistent apparent vertical line in the bottom subplot of Fig.~\ref{Fig: prepfold}c) is only stable in the demodulated time series at a very specific companion mass between 0.9867 $< q <$ 0.9899.
For example, the $q$ value behind Fig.~\ref{Fig: prepfold}c is 0.98767, leading to a PRESTO significance level of $P(\text{Noise})<2\times10^{-15}\, $ or $7.9\sigma$.
This limited mass-ratio range is a strong suggestion for an astrophysical origin, which is hard to mimic. 
However, no pulsar (or RFI) is visible in the frequency-phase diagram (Fig.~\ref{Fig: Cand-42ms}) composed by dedispersing the original, RFI-cleaned filterbank file at full time and frequency resolution, and folding it with the ephemeris expected for the candidate orbit, given the trial mass ratio $q$.
Furthermore, the candidate is not found at any other epoch. 
Therefore, we conclude that the visual notability of this candidate is due to a chance alignment of noise outliers in the low-resolution time-frequency plot (Fig.~\ref{Fig: prepfold}c).
This diagnostic plot is composed of 64 profile bins by 64 sub-integrations.
The number of trials is much larger than this, as it is spanned by the search ranges in over period, period derivative, and mass ratio $q$. Based on experience in previous pulsar surveys \citep[e.g.,][]{Coenen14, vL2020}, such 8$\sigma$ chance outliers are not uncommon, and are hardly ever confirmed.

\begin{figure*}
    \centering\includegraphics[width=0.8\linewidth]{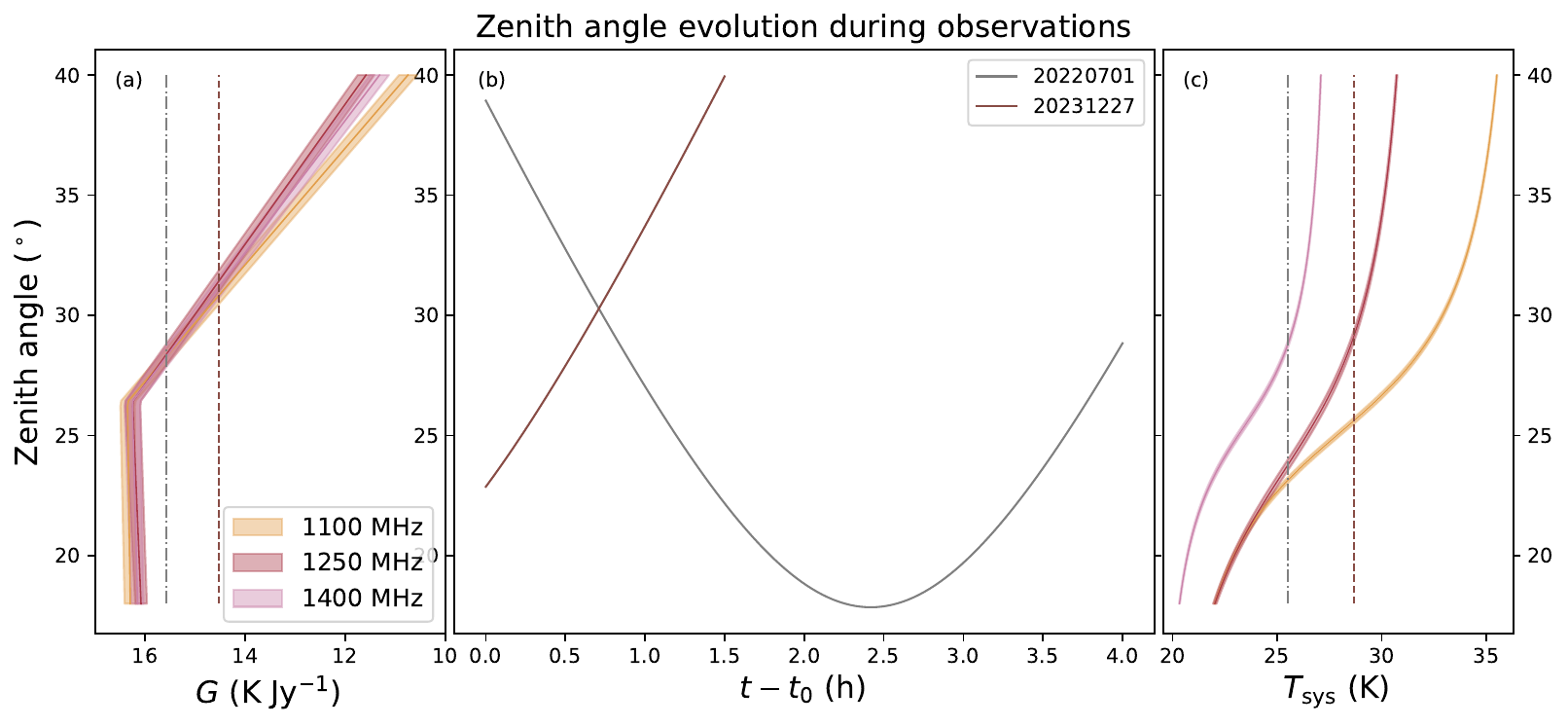}
    \caption{FAST observation and performance evaluation for our observations. 
    Panel (a) and (c) delineate the relationship between telescope gain $G$, system temperature $T_\mathrm{sys}$ and the zenith angle for the central beam (M01) at 1100\,MHz, 1250\,MHz and 1400\,MHz respectively.
    The middle panel (b) shows the zenith angle evolution during our 20220701 4-hr and 20231227 1.5-hr observations. 
    The estimated mean $G$ and $T_\mathrm{sys}$ are indicated with vertical lines in (a) and (c). FAST parameters taken from \citet{Jiang2020}.
    The evolution of the gain in panel (a) is because the aperture efficiency is mostly unchanged until a zenith angle of 26.4$^\circ$, then starts to decrease.
    For zenith angles larger than 15$^\circ$, local sidelobe RFI increases $T_\mathrm{sys}$ (c).}
    \label{Fig: zenith-angle}
\end{figure*}

\begin{figure}
    \centering\includegraphics[width=1.0\linewidth]{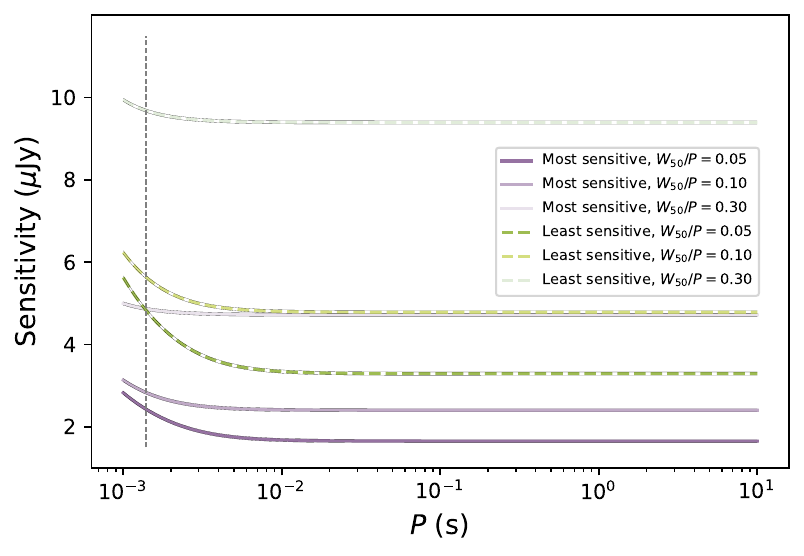}
    \caption{Search sensitivity limit as a function of the pulsar spin period $P$. We take the intrinsic $W_{50}/P$ to be 0.05, 0.10 and 0.30; the instrumental broadening and DM smearing are included when calculating the effective pulse width. The vertical dashed line is the fastest-spinning pulsar ever found.}
    \label{Fig: sensitivity}
\end{figure}

\begin{figure}
    \centering\includegraphics[width=1.0\linewidth]{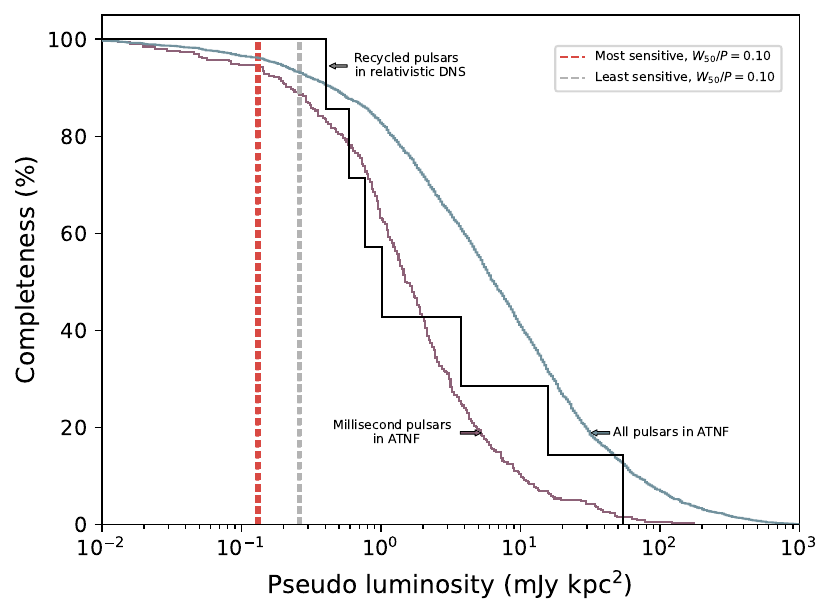}
    \caption{Search completeness as a function of the pseudo luminosity $L_\mathrm{pseudo}=S_\mathrm{min} d^2$. The cumulative distribution of $L_\mathrm{pseudo}$ for three (sub)sets of pulsars in the ATNF catalog are shown:
    all (blue), MSPs ($P<20$\,ms) in purple, and recycled pulsars in relativistic DNS systems (black). To estimate the $L_\mathrm{pseudo}$ of the companion, we used the HI-absorption derived distance (7.4\,kpc), and we assumed an intrinsic $W_{50}/P$ of 0.10, and we considered both the most (red) and least (gray) sensitive observations.
    For recycled pulsars in relativistic DNS systems, our search is 100\% complete; while for the general pulsar population, our search completeness exceeds 90\%.}
    \label{Fig: completeness}
\end{figure}

\section{Discussion}

Below we discuss the upper limits we obtain (Sect.~\ref{sec:limit}), the chances that the companion is a white dwarf or radio-quiet neutron star (Sect.~\ref{sec:expl}, \ref{sec:ppd}), and the forecast for when a companion pulsar may become visible (Sect.~\ref{sec:prob}, \ref{sec:fb}).

\subsection{Upper limit on the companion flux density}
\label{sec:limit}
To estimate the upper limit on the source flux density (the minimum detectable flux density $S_{\min}$), we used the standard radiometer equation for periodic, narrow signals \citep[following, e.g.,][]{ls10},
\begin{equation}
    S_{\min} = \beta_\mathrm{DF} \frac{\left(S / N_{\min}\right) T_{\mathrm{sys}}}{G \sqrt{n_{\mathrm{p}} \Delta \nu t_{\mathrm{int}}}} \sqrt{\frac{W}{P-W}},
\label{eq:smin}
\end{equation}
where $\beta_\mathrm{DF}$\footnote{Subscript added to distinguish from the impact parameter $\beta$, used later.} is the telescope degradation factor, S/N$_{\min}$ is the pulsar detection S/N threshold, $T_\mathrm{sys} = T_\mathrm{rec} + T_\mathrm{sky}$ is the overall system temperature, $G$ is the telescope gain, $n_\mathrm{p}$ is the number of polarization, $\Delta\nu$ is the observing frequency bandwidth, $t_\mathrm{int}$ is the integration time, and $W$ and $P$ are the pulse width and the pulsar spin period respectively. The effective pulse width $W$ of a radio pulsar is broadened due to several effects \citep{Mikhailov2017}:
\begin{equation}
W = \sqrt{\langle W_{50}/P\rangle^{2} P^2+t_{\mathrm{scat}}^{2}+t_{\mathrm{DM}}^{2}+t_{\mathrm{samp}}^{2}},
\label{eq:weff}
\end{equation}
where $W_{50}/P$ is the pulsar duty cycle (isolated here because it varies less than $W_{50}$ itself), $t_{\mathrm{scat}} \le 0.01$\,ms is the scattering time, while $t_{\mathrm{samp}} = 0.049152$\,ms and $t_{\mathrm{DM}} \simeq 0.11293$\,ms are the sampling time broadening and dispersion smearing respectively.
For cases where $W \gtrsim P$, Eq.~\ref{eq:smin} is no longer valid. Given the relatively high observing frequency and moderate DM, instrumental effects are unlikely to cause the effective width to exceed the period, and we discuss the influence of the intrinsic duty cycle on our search sensitivity below.
We remain sensitive to any such companions with a duty cycle larger than unity, which would appear as one, diminished harmonic in the power spectrum, as we also search for single-harmonic candidates (Sect.~\ref{sec:datareduc}.)

For FAST, the degradation factor $\beta_\mathrm{DF} \simeq 1.5$ and the total bandwidth $\Delta\nu \simeq 400$\,MHz (after taking into account the channels excluded by RFI removal). 
For the most sensitive observation, we took the 4-hour observation on 1 July 2022 as an example. The zenith angle during this observation varies between 20$^\circ$ and 40$^\circ$. 
Based on \citet{Jiang2020} we estimated the average $G \simeq 15.6$\,K\,Jy$^{-1}$ and $T_\mathrm{sys} \simeq 25.5$\,K. Consequently, for S/N$_\mathrm{min} = 10$ and a typical MSP duty cycle of $W_{50}/P$ = 0.10 (following \citealt{Mikhailov2017}), we find $S_\mathrm{min} \simeq 2.4\times 10^{-6}$\,Jy.

Figure~\ref{Fig: zenith-angle} shows the system sensitivity evolution for both this most sensitive observation and the least sensitive one on 27 December 2023.

Figure~\ref{Fig: sensitivity} next shows our search sensitivity limit $S_\mathrm{min}$ as a function of spin period $P$, for the most and least sensitive cases, and for $W_{50}/P$ values of 0.05, 0.10 and 0.30. 
It is visible there that $S_\mathrm{min}$ increases when the pulsar spins very fast, because the other factors in Eq.~\ref{eq:weff} start to dominate, effectively smearing out the pulse profile.
The fastest-spinning neutron star observed up to now is PSR J1748-2446ad, which has a spin frequency of 716 Hz \citep{Hessels2006}. One theoretical pulsar spin frequency limit is given by $f_\mathrm{max} \simeq 1045\,(M/M_\odot)^{1/2}(10\,\mathrm{km}/R)^{3/2}$\,Hz \citep{Lattimer2004}. 
For the companion of \psr, $M_c = 1.322$\,\msun and $R$ is most likely in the range 10$-$14\,km \citep{Ozel2016}. Hence, we have a $f_\mathrm{max} \simeq 1202$\,Hz (or period 0.83\,ms). 
This is, however, a mostly theoretical limit in our case. Binary evolution theory suggests that the companion that we search for should be an only mildly recycled pulsar, with a spin period most likely between 10\,ms and 100\,ms. Therefore, only the flat part of the $S_\mathrm{min}$$-$$P$ curves in Fig.~\ref{Fig: sensitivity} are used in further discussions. 

To place these limits in perspective, we converted them to a pseudo luminosity, which we can then compare against the known pulsar (sub)populations. 
This pseudo luminosity $L_\mathrm{pseudo} = S_\mathrm{min} d^2$ depends on the pulsar distance $d$, determined to be $7.4^{+2.5}_{-1.4}$\,kpc from HI absorption (\citealt{vanLeeuwen2015}; the earlier, DM-based distance was $\sim$5.4\,kpc.)
The search completeness as a function of $L_\mathrm{pseudo}$ is shown in Fig.~\ref{Fig: completeness}.
The cumulative distribution for all pulsars, all MSPs and eight recycled pulsars in relativistic DNSs from ATNF are plotted in blue, purple and black curves, respectively. We also displayed the most and least sensitive cases in our search with red and gray vertical lines for different $d$.

For recycled pulsars in relativistic DNSs, the type of companion we expect, our search completeness $C_\mathrm{search}$~(\%) reaches 100\%; i.e., even at this distance we would have detected every known recycled pulsar in a relativistic DNSs. 
The general pulsar population also contains some faint pulsars with a $L_\mathrm{pseudo}$ below 0.1\,mJy\,kpc$^2$; even the general population we reach a $C_\mathrm{search}$~(\%) $\simeq$ 94\%.

Based on these considerations, we conclude that no periodic pulsar signals are present in our data.

\subsection{Possible explanations for the non-detection of the companion signal}
\label{sec:expl}
Our non-detection suggests, again, that the companion is either not a neutron star (but a massive WD instead); 
or a neutron star that does not emit radio emission; or a pulsar whose beam misses our LOS. The statistics of the last option are discussed in detail in the next section. 

In the first option listed above, the \psr system would contain a young pulsar with an older WD companion.
As the most massive component of the original binary star system explodes in a supernova first, and normally produces the most massive compact object, one generally expects the most massive component in a relativistic binary to be the oldest. 

And yet PSR J1141$-$6545, a young, relativistic binary pulsar, is in a $\sim$ 4.74 hr eccentric orbit with an older 1.08\,\msun WD companion \citep{VenkatramanKrishnan2019}. 
The companion was identified as a WD through its detection in the optical \citep{Antoniadis2011}. The temporal evolution of the orbital inclination of this pulsar indicates the WD is rapidly rotating \citep{VenkatramanKrishnan2020}. 
The companion of \psr is significantly more massive, $m_\mathrm{c} = 1.32\,\msun$, and hence more likely to be a neutron star. 
That said, PSR B2303+46 is another young pulsar and old WD system \citep{vanKerkwijk1999}; here the WD mass is
1.3\,\msun. Based on mass alone, the companion can possibly still be an exceptionally massive WD.
For a mildly recycled neutron star the companion mass is well within the expected range \citep{vanLeeuwen2015} so this remains more likely.

The eccentricity of \psr ($e=0.085$) is lower than that of the pulsar-WD systems PSR J1141$-$6545 ($e=0.172$) and PSR
B2303+46 ($e=0.658$) but closer to the double pulsar PSR J0737$-$3039A ($e=0.088$).
Admittedly, this sample contains only few systems, and the binary evolution history for these sources likely differs somewhat, too \citep[see, e.g.,][]{Davies2002}.

\begin{figure}
    \centering
    \includegraphics[width=1.0\linewidth]{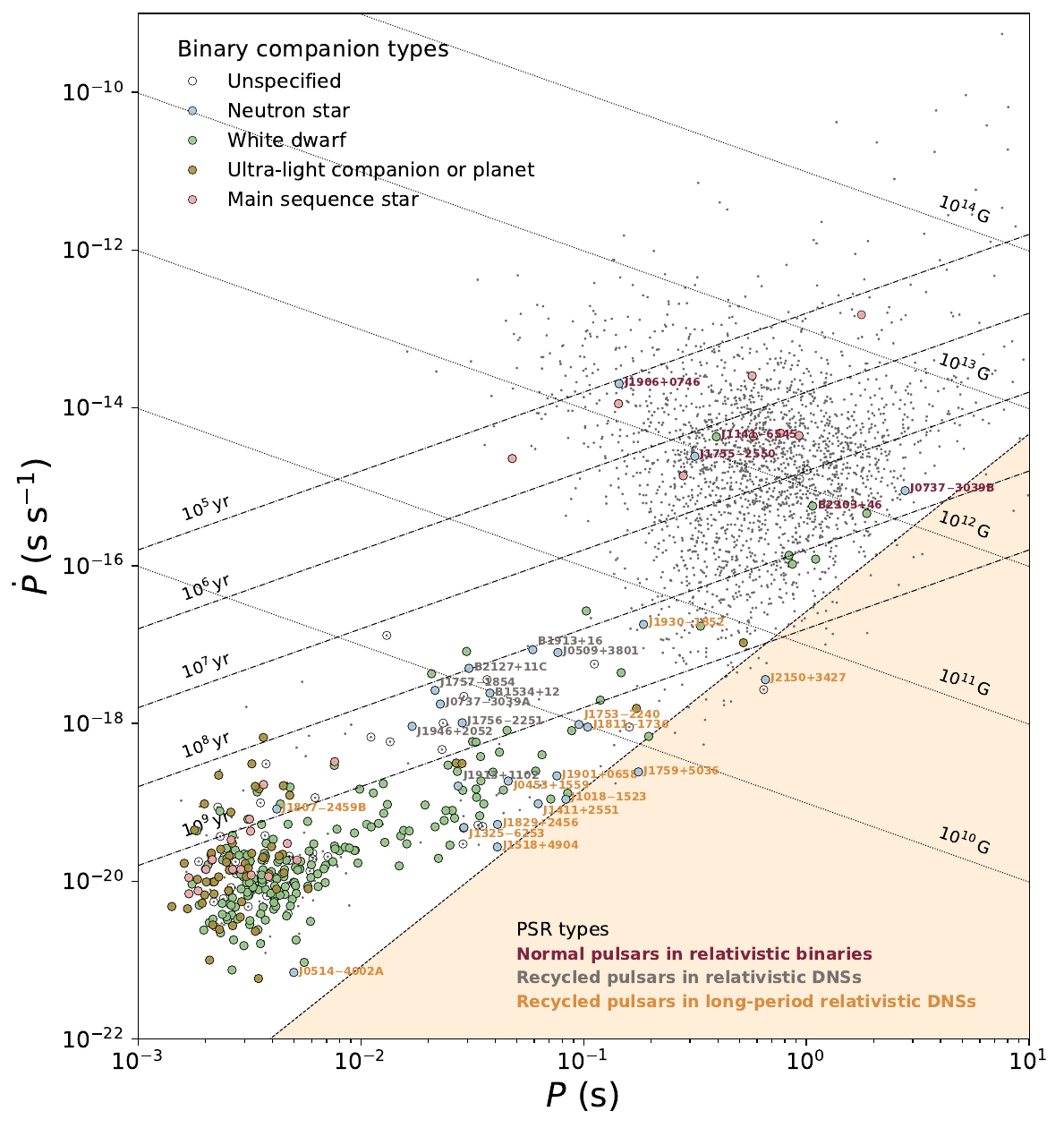}
    \caption{$P$-$\dot{P}$ diagram that shows the distribution of different types of companions in the binaries. The known pulsars in relativistic and/or DNS systems are labeled in text with different colors. Made using \texttt{psrqpy} \citep{psrqpy} based on the Australia Telescope National Facility (ATNF) pulsar catalog\protect\footnotemark ~\citep{Manchester2005}. }
    \label{Fig: P-Pdot}
\end{figure}

In the second option listed above, the companion is a neutron star that does not emit at all.
This could be because it was mildly recycled, but then spun down until it crossed the death line
(cf. Fig.~\ref{Fig: P-Pdot}; also \citealt{lv04}) again.
This would signify a large delay between the formation of both pulsars (\psr is very young, after all, while the companion would need to be exceedingly old, >1\,Gyr), there is some evidence supporting this in the $P$-$\dot{P}$ diagram (Fig.~\ref{Fig: P-Pdot}).
A number of mildly recycled pulsars (such as PSR~J1518+4904, J1018$-$1523 and J1759+5036) have reached the death line, and there are none found beyond. This shows that this subpopulation, in time, also switches off. 

Other ``radio-quiet'' neutron stars have been hypothesized to never emit in radio at all \citep[see][and references therein]{Pastor2023}. This group, however, can not be very large, as there arguably are already more radio-bright neutron stars than the Galactic supernova rate can provide.

We conclude that it is possible that the companion is a neutron star that does not emit at all. 
More likely, though, it does, but beamed away from us.
We pursue this study more later in this section, after a brief aside on the $P$-$\dot{P}$ diagram.

\footnotetext{\url{https://www.atnf.csiro.au/research/pulsar/psrcat/}}

\subsection{Further insights from the $P$-$\dot{P}$ diagram}
\label{sec:ppd}

In the $P-\dot{P}$ diagram (Fig.~\ref{Fig: P-Pdot}), the young, normal pulsars in relativistic binaries stand out in the top right (marked with red names). 
These are the systems mentioned in Sect.~\ref{sec:expl}, plus PSR~J1755$-$2550, a relatively young pulsar with a similar eccentricity ($e\sim 0.089$) as \psr but a rather uncertain companion mass between 0.4 and 2.0\,\msun \citep{Ng2018}.
These sources are not recycled, and their active lifetimes are shorter than those of MSPs.
Of these, \psr is the youngest.

To allow us to focus our discussion on the relevant pulsars and companions, we follow the ATNF catalog \citep{Manchester2005} and categorized the binary companions in this diagram into five species: neutron star, white dwarf, ultra-light companion or planet, and main sequence star.
The known pulsars in relativistic and/or DNS systems (see \cite{vanLeeuwen2015} and references therein) are labeled by text of different colors. 

The neutron-star companion sources cluster between 20$-$100\,ms (Fig.~\ref{Fig: P-Pdot}, blue markers).
As this is where we expect a pulsar companion to \psr too, we have paid special attention to this period range (see Sect.~\ref{sec:cands:42}). We did not, however, detect a pulsar in our search. 
One of the best-known and most studied systems in this cluster is the Hulse-Taylor binary, PSR~B1913+16.
In that system, the recycled pulsar is seen, but searches for the unrecycled, younger pulsar were unsuccessful \citep[see, e.g.,][]{Taylor76}. 
In that sense, the visibility of the components in PSR~B1913+16 is the exact opposite of {\psr}.
The non-detection of the companion to PSR~B1913+16 can be explained by the same reasons that hold for a non-detected neutron star companion to {\psr}; but as the unrecycled companion to PSR~B1913+16 is expected to be relatively short-lived, it is especially likely that it has passed the death line.

\subsection{Probability of companion visibility, and constraints on viewing geometry}
\label{sec:prob}

\begin{figure}
    \centering
    \subfigure{\includegraphics[width=1.0\linewidth,trim={0 0.3cm 0 0},clip]{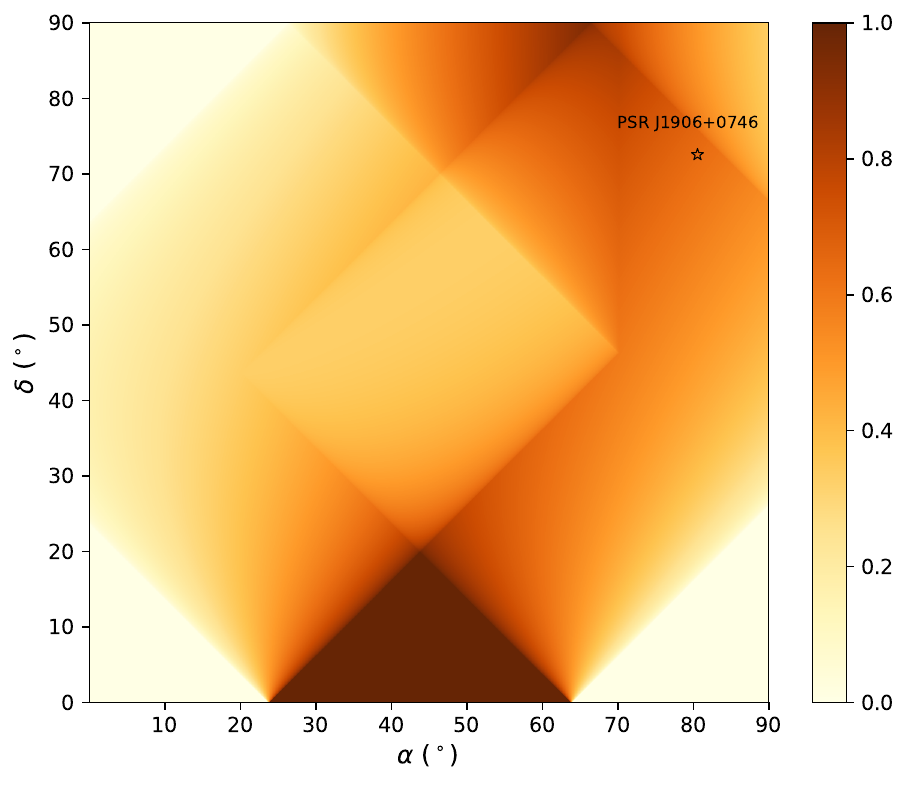}}
    \subfigure{\includegraphics[width=1.0\linewidth,trim={0 0 0 0.2cm},clip]{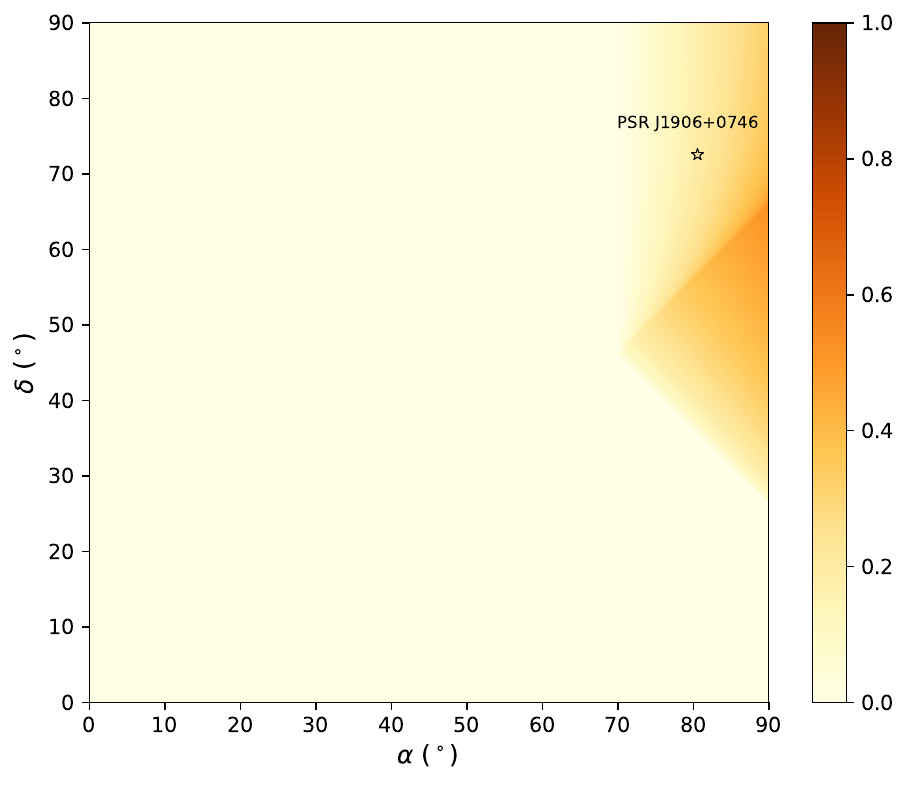}}
    \caption{Probability of seeing the companion at a single arbitrary time as a function of magnetic inclination angle $\alpha$ and spin-orbit misalignment $\delta$. 
    The orbital inclination angle $i$ is fixed to 43.7$^\circ$. Both $\alpha$ and $\delta$ are symmetrical about 90$^\circ$; therefore, we only show the [0$^\circ$, 90$^\circ$] $\times$ [0$^\circ$, 90$^\circ$] quadrant.
    \psr itself is marked with a star.
    \textit{Upper panel}: probability of observing emission from at least one pole; \textit{bottom panel}: probability of observing emission from both poles. The assumptions here are: (1) the beam extent in the latitudinal direction is 40$^\circ$ for both MP and IP (equivalently, the impact parameter $\beta$ is [$-$20$^\circ$, 20$^\circ$]); (2) we are observing using an ideal telescope and the probability is from purely geometrical consideration. For \psr, ($\alpha$, $\delta$) is (80.54$^\circ$, 72.55$^\circ$), from Wang et al.~(2025, in prep.).
    }
    \label{Fig: Observing-probability}
\end{figure}

We estimated the chance that, at a certain single moment, the companion is visible from Earth, as a function of magnetic inclination angles $\alpha$ and spin-orbit misalignment $\delta$. The probability is calculated as the fraction of one precession cycle that the beam points at us. Here we assumed the total beam size in the latitudinal direction is 40$^\circ$ for both MP and IP (as in \citealt{Desvignes2019} and Wang et al.~2025, in prep.) and that we are observing using an ideal telescope (i.e., that we can always detect the pulsar when our LOS cuts through the beam). 
The probability contours for observing both poles and at least one pole, are shown in Fig.~\ref{Fig: Observing-probability}. For reference, we also marked \psr itself with a star.

When $\delta=0^\circ$, there is no practical effect from the geodetic precession; in this case, we can either always see or never see the companion, depending on $\alpha$.
In general, when the second neutron star is formed, the mass loss and the kick associated with the supernova explosion will usually cause a misalignment between the spin axis of the existing, recycled pulsar and the orbital angular momentum \citep{Tauris2017}. If this misalignment was significant (for example, $\delta \ge 20^\circ$), there is still a chance that we will see the recycled companion to \psr in the future.
A range of misalignment angles have been measured or suggested for similar systems.
In one such system, PSR~J0737$-$3039A, the angle it is actually low, 3.2$^\circ$ \citep{Ferdman2013}, suggesting a symmetric supernova or one imparting a kick opposite to its pre-explosion orbital velocity \citep{Willems2004}.
On the other hand \citet{Fonseca2014} find the preferred angle for PSR~B1534+12 is 27$\pm$3{\degr}.
Some subtle selection effect may be at play though, as in these cases the recycled pulsar was discovered first, while in our case, the young pulsar was discovered first.

\begin{figure}
    \centering
    \includegraphics[width=1.0\linewidth]{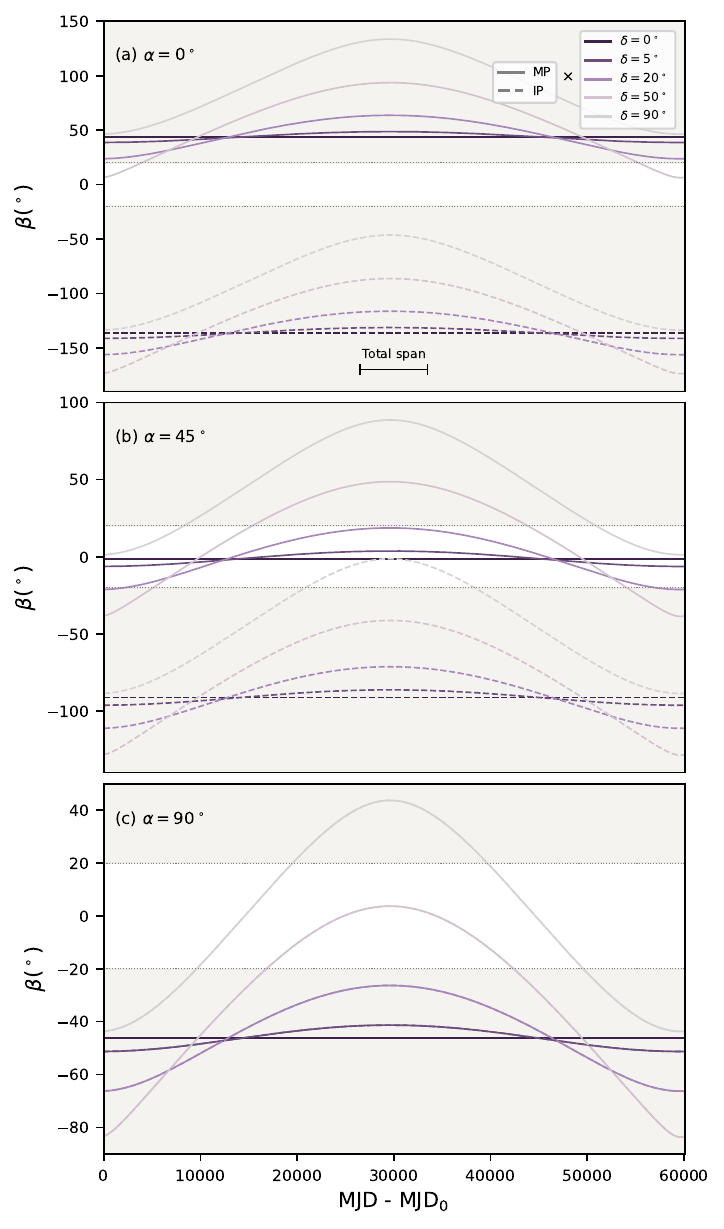}
    \caption{The evolution of the impact angle  $\beta$
    in one precession cycle for the putative companion pulsar for different magnetic inclination angles $\alpha$ and spin-orbit angles $\delta$ ($0^\circ$, $5^\circ$, $20^\circ$, $50^\circ$ and $90^\circ$). 
    In panel (a), $\alpha = 0^\circ$panel (b) $45^\circ$ and panel (c) $90^\circ$. For (a) and (b), the $\beta_\mathrm{MP}$ and $\beta_\mathrm{IP}$ are plotted using solid and dashed lines, respectively; for (c), $\beta_\mathrm{MP}$ and $\beta_\mathrm{IP}$ curves coincide with each other. 
    The invisible $\beta$ regions are filled with gray shadow. 
    Since the reference precessional orbital phase $\Phi_0$ of the companion is unknown, the reference MJD$_0$ is chosen to be the epoch that $\Phi_0 = 0^\circ$.
    For reference, a scale bar that shows the observing duration of \psr ($\sim 7000$\,days) is plotted at the bottom of panel (a). This span can occur anywhere in the cycle. }
    \label{fig: beta_and_geometry}
\end{figure}

The precession rate of the presumed companion pulsar is $\sim$\,2.19$^\circ$\,yr$^{-1}$. Although the companion spin-orbit misalignment angle $\delta$ is unknown, we can qualitatively constrain a number of general $\delta$ and $\alpha$ scenarios, based on the non-detections between 2006 and 2024.
We calculated the evolution of the impact parameter $\beta$ (the angle between the magnetic axis and the LOS) for the companion, for different $\delta$ ($0^\circ$, $5^\circ$, $20^\circ$, $50^\circ$, and $90^\circ$) and different $\alpha$ ($0^\circ$, $45^\circ$ and $90^\circ$) values, using the \citet{Kramer2009} precessional model.
The results are shown in Fig.~\ref{fig: beta_and_geometry}.
This provides constraints on the viewing geometry if the companion is indeed a pulsar:
when $\alpha$ is very small, $\delta$ is unconstrained (Fig.~\ref{fig: beta_and_geometry}a); a moderate $\alpha$ (e.g. 45$^\circ$) favors a large $\delta$ ($\gtrsim 50^\circ$), visible in Fig.~\ref{fig: beta_and_geometry}b);
while if the viewing geometry is nearly orthogonal, the current time span ($\sim$ 7000 days) is not yet quite enough to constrain $\delta$ (Fig.~\ref{fig: beta_and_geometry}c).

While this misalignment angle $\delta$ of the invisible companion is hard to determine a priori, some information can in principle be inferred from the eccentricity of the system, a quantity that can be determined from the visible binary component.
That is because both the eccentricity and the misalignment angle are in large part determined by the supernova kick that created the second neutron star.
In the double pulsar PSR J0737$-$3039A, for example, both are low ($e=0.088$ and $\delta=3.2^\circ$) while in PSR~B1534+12 both are higher ($e=0.27$ and $\delta=27^\circ$). 
The low eccentricity of \psr of $e=0.085$ thus probably also influences the prior on its companion misalignment angle $\delta$, by some fraction that is currently unknown to us.

\subsection{All-time beaming fraction of a precessing pulsar}
\label{sec:fb}

The pulsar beaming fraction (the portion of the unit sky over which a pulsar is visible) has been studied in the past, especially in the context of pulsar energetics and population studies.
This fraction was, however, not studied before for a precessing pulsar.
As the spin axis tilts, different parts of the sky are illuminated by the pulsar beam.

\begin{figure}
    \centering
    \includegraphics[width=1.0\linewidth]{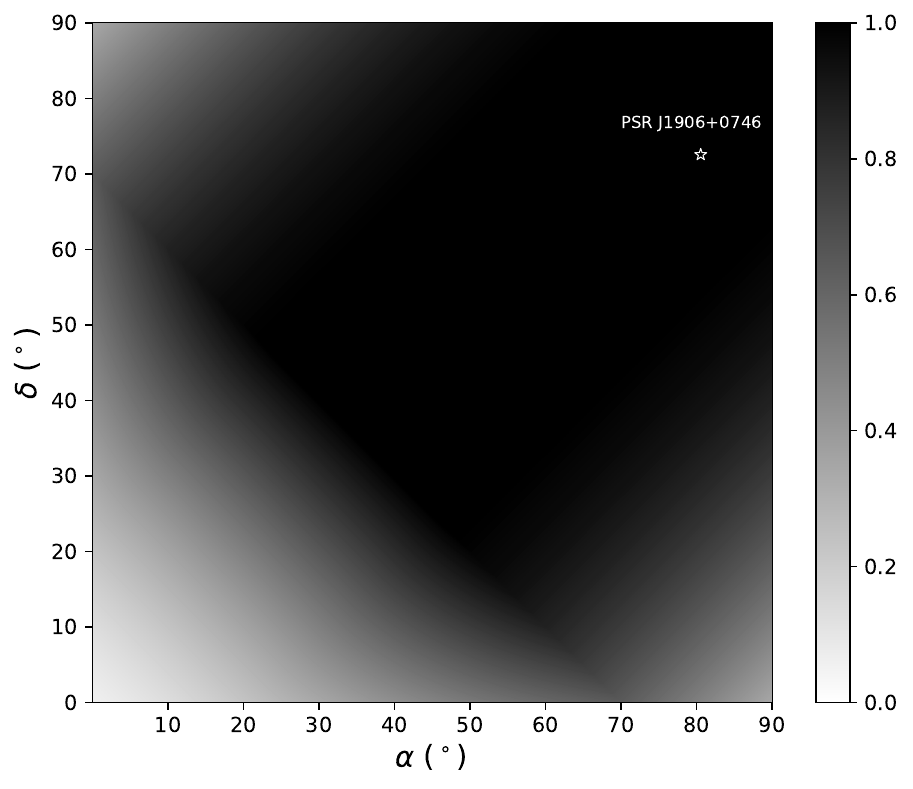}
    \caption{The all-time beaming fraction $f_\mathrm{b, tot}$ of a precessing pulsar as a function of magnetic inclination angle $\alpha$ and spin-orbit misalignment $\delta$. Here, the latitudinal beam radius $\rho$ is 20$^\circ$. The location of \psr in the $\alpha$-$\delta$ space is marked as in Fig.~\ref{Fig: Observing-probability}.}
    \label{fig: beaming_fraction}
\end{figure}

Here, we considered a pulsar with magnetic inclination angle $\alpha$, misalignment $\delta$, and a paddle beam with latitudinal radius $\rho$ (cf.~Wang et al.~2025, in prep.).
The all-time beaming fraction $f_\mathrm{b, tot}$ represents the fraction of the sky from which the precessing pulsar is visible at some point.
It is the union of all instantaneous beaming fractions $f_\mathrm{b}(t)$ at times $t$. 
This $f_\mathrm{b, tot}$ is given by $(\sin\theta_2 - \sin\theta_1)/2$, where $\theta_1$, $\theta_2$ are the lower and upper latitudes of the illuminated strip in the orbital plane frame (where the $\vec{L_\mathrm{tot}}$ aligns with the $z$ direction), respectively.
Figure~\ref{fig: beaming_fraction} shows the values of $f_\mathrm{b, tot}$. 

\psr has an all-time sky coverage of 1.
This means that during one precession cycle, it will not only sweep over Earth (as we know), but over all potential other observers too. It is thus discoverable from all directions, at some point. 
It is apparent that $\alpha$ and $\delta$ are the determining factors for the all-time beaming fraction $f_\mathrm{b, tot}$: if a pulsar has a large combination of $\alpha+\delta$, it can illuminate almost all the sky during one precessing cycle, in $O(10^2)$ years for typical DNS systems; this strongly suggests that if we keep observing, we will eventually discover all these currently hidden pulsars.

\section{Conclusion}

In $>$2 years of monthly, highly sensitive FAST observation we did not detect pulsations from the companion of {\psr}.
The mass of the companion fits a recycled neutron star best, but a massive white dwarf is not ruled out.
This companion neutron star most likely emits in radio, but its beam does not sweep across us.
We find that for most combinations of the spin-orbit misalignment (very probably present, given the supernova kick likely imparted on \psr), and the magnetic inclination angle, we may well observe the companion pulsar at some time in the future.
We conclude it remains highly meaningful to extend the search we presented over a much longer time span. 

\begin{acknowledgement}
We thank Alessandro Ridolfi for sharing \texttt{pysolator}.
We thank Chenchen Miao and Pei Wang for assistance with the FAST data transfer, and 
Gregory Desvignes, Ingrid Stairs, Laila Vleeschower Calas, Ben Stappers, Michael Kramer, Di Li, and Weiwei Zhu for their support at various proposal stages.
This research was supported by Vici research project ``ARGO'' (grant number 639.043.815), and through CORTEX (NWA.1160.18.316), under the research programme NWA-ORC; both financed by the Dutch Research Council (NWO).
This work has used the data from the Five-hundred-meter Aperture Spherical radio Telescope (FAST) under proposals PT2021\_0001, PT2022\_0024 and PT2023\_0010.
FAST is a Chinese national mega-science facility, operated by the National Astronomical Observatories of Chinese Academy of Sciences (NAOC). 

\end{acknowledgement}

\bibliographystyle{yahapj}
\bibliography{export-bibtex}

\begin{thebibliography}{}
\providecommand\natexlab[1]{#1}
\providecommand\JournalTitle[1]{#1}

\bibitem[{{Antoniadis} {et~al.}(2011){Antoniadis}, {Bassa}, {Wex}, {Kramer}, \& {Napiwotzki}}]{Antoniadis2011}
{Antoniadis}, J., {Bassa}, C.~G., {Wex}, N., {Kramer}, M., \& {Napiwotzki}, R. 2011, \href{http://dx.doi.org/10.1111/j.1365-2966.2010.17929.x}{\JournalTitle{\mnras}, 412, 580}

\bibitem[{{Bilous} {et~al.}(2015){Bilous}, {Pennucci}, {Demorest}, \& {Ransom}}]{Bilous2015}
{Bilous}, A.~V., {Pennucci}, T.~T., {Demorest}, P., \& {Ransom}, S.~M. 2015, \href{http://dx.doi.org/10.1088/0004-637X/803/2/83}{\JournalTitle{\apj}, 803, 83}

\bibitem[{{Burgay} {et~al.}(2003){Burgay}, {D'Amico}, {Possenti}, {Manchester}, {Lyne}, {Joshi}, {McLaughlin}, {Kramer}, {Sarkissian}, {Camilo}, {Kalogera}, {Kim}, \& {Lorimer}}]{Burgay2003}
{Burgay}, M., {D'Amico}, N., {Possenti}, A., {et~al.} 2003, \href{http://dx.doi.org/10.1038/nature02124}{\JournalTitle{\nat}, 426, 531}

\bibitem[{{Coenen} {et~al.}(2014){Coenen}, {van Leeuwen}, {Hessels}, {Stappers}, {Kondratiev}, {Alexov}, {Breton}, {Bilous}, {Cooper}, {Falcke}, {Fallows}, {Gajjar}, {Grie{\ss}meier}, {Hassall}, {Karastergiou}, {Keane}, {Kramer}, {Kuniyoshi}, {Noutsos}, {Os{\l}owski}, {Pilia}, {Serylak}, {Schrijvers}, {Sobey}, {ter Veen}, {Verbiest}, {Weltevrede}, {Wijnholds}, {Zagkouris}, {van Amesfoort}, {Anderson}, {Asgekar}, {Avruch}, {Bell}, {Bentum}, {Bernardi}, {Best}, {Bonafede}, {Breitling}, {Broderick}, {Br{\"u}ggen}, {Butcher}, {Ciardi}, {Corstanje}, {Deller}, {Duscha}, {Eisl{\"o}ffel}, {Fender}, {Ferrari}, {Frieswijk}, {Garrett}, {de Gasperin}, {de Geus}, {Gunst}, {Hamaker}, {Heald}, {Hoeft}, {van der Horst}, {Juette}, {Kuper}, {Law}, {Mann}, {McFadden}, {McKay-Bukowski}, {McKean}, {Munk}, {Orru}, {Paas}, {Pandey-Pommier}, {Polatidis}, {Reich}, {Renting}, {R{\"o}ttgering}, {Rowlinson}, {Scaife}, {Schwarz}, {Sluman}, {Smirnov}, {Swinbank}, {Tagger}, {Tang}, {Tasse}, {Thoudam}, {Toribio}, {Vermeulen}, {Vocks}, {van
  Weeren}, {Wucknitz}, {Zarka}, \& {Zensus}}]{Coenen14}
{Coenen}, T., {van Leeuwen}, J., {Hessels}, J. W.~T., {et~al.} 2014, \href{http://dx.doi.org/10.1051/0004-6361/201424495}{\JournalTitle{\aap}, 570, A60}

\bibitem[{Damour \& Ruffini(1974)}]{dr74}
Damour, T. \& Ruffini, R. 1974, \href{https://ui.adsabs.harvard.edu/#abs/1974CRASM.279..971D/abstract}{\JournalTitle{Academie des Sciences Paris Comptes Rendus Ser.\,Scie.\,Math.}, 279, 971}

\bibitem[{{Davies} {et~al.}(2002){Davies}, {Ritter}, \& {King}}]{Davies2002}
{Davies}, M.~B., {Ritter}, H., \& {King}, A. 2002, \href{http://dx.doi.org/10.1046/j.1365-8711.2002.05594.x}{\JournalTitle{\mnras}, 335, 369}

\bibitem[{{Desvignes} {et~al.}(2019){Desvignes}, {Kramer}, {Lee}, {van Leeuwen}, {Stairs}, {Jessner}, {Cognard}, {Kasian}, {Lyne}, \& {Stappers}}]{Desvignes2019}
{Desvignes}, G., {Kramer}, M., {Lee}, K., {et~al.} 2019, \href{http://dx.doi.org/10.1126/science.aav7272}{\JournalTitle{Science}, 365, 1013}

\bibitem[{{Ferdman} {et~al.}(2013){Ferdman}, {Stairs}, {Kramer}, {Breton}, {McLaughlin}, {Freire}, {Possenti}, {Stappers}, {Kaspi}, {Manchester}, \& {Lyne}}]{Ferdman2013}
{Ferdman}, R.~D., {Stairs}, I.~H., {Kramer}, M., {et~al.} 2013, \href{http://dx.doi.org/10.1088/0004-637X/767/1/85}{\JournalTitle{\apj}, 767, 85}

\bibitem[{{Fonseca} {et~al.}(2014){Fonseca}, {Stairs}, \& {Thorsett}}]{Fonseca2014}
{Fonseca}, E., {Stairs}, I.~H., \& {Thorsett}, S.~E. 2014, \href{http://dx.doi.org/10.1088/0004-637X/787/1/82}{\JournalTitle{\apj}, 787, 82}

\bibitem[{{Hessels} {et~al.}(2006){Hessels}, {Ransom}, {Stairs}, {Freire}, {Kaspi}, \& {Camilo}}]{Hessels2006}
{Hessels}, J. W.~T., {Ransom}, S.~M., {Stairs}, I.~H., {et~al.} 2006, \href{http://dx.doi.org/10.1126/science.1123430}{\JournalTitle{Science}, 311, 1901}

\bibitem[{{Jiang} {et~al.}(2020){Jiang}, {Tang}, {Hou}, {Liu}, {Kr{\v{c}}o}, {Qian}, {Sun}, {Ching}, {Liu}, {Duan}, {Yue}, {Gan}, {Yao}, {Li}, {Pan}, {Yu}, {Liu}, {Li}, {Peng}, {Yan}, \& {FAST Collaboration}}]{Jiang2020}
{Jiang}, P., {Tang}, N.-Y., {Hou}, L.-G., {et~al.} 2020, \href{http://dx.doi.org/10.1088/1674-4527/20/5/64}{\JournalTitle{Research in Astronomy and Astrophysics}, 20, 064}

\bibitem[{{Kargaltsev} \& {Pavlov}(2009)}]{Kargaltsev2009}
{Kargaltsev}, O. \& {Pavlov}, G.~G. 2009, \href{http://dx.doi.org/10.1088/0004-637X/702/1/433}{\JournalTitle{\apj}, 702, 433}

\bibitem[{{Kasian}(2012)}]{kasi11}
{Kasian}, L. 2012, PhD thesis, University of British Columbia, {\url{https://circle.ubc.ca/handle/2429/41515}}

\bibitem[{{Kramer} \& {Stairs}(2008)}]{Kramer2008}
{Kramer}, M. \& {Stairs}, I.~H. 2008, \href{http://dx.doi.org/10.1146/annurev.astro.46.060407.145247}{\JournalTitle{\araa}, 46, 541}

\bibitem[{{Kramer} \& {Wex}(2009)}]{Kramer2009}
{Kramer}, M. \& {Wex}, N. 2009, \href{http://dx.doi.org/10.1088/0264-9381/26/7/073001}{\JournalTitle{Classical and Quantum Gravity}, 26, 073001}

\bibitem[{{Kramer} {et~al.}(2006){Kramer}, {Stairs}, {Manchester}, {McLaughlin}, {Lyne}, {Ferdman}, {Burgay}, {Lorimer}, {Possenti}, {D'Amico}, {Sarkissian}, {Hobbs}, {Reynolds}, {Freire}, \& {Camilo}}]{Kramer2006}
{Kramer}, M., {Stairs}, I.~H., {Manchester}, R.~N., {et~al.} 2006, \href{http://dx.doi.org/10.1126/science.1132305}{\JournalTitle{Science}, 314, 97}

\bibitem[{{Lattimer} \& {Prakash}(2004)}]{Lattimer2004}
{Lattimer}, J.~M. \& {Prakash}, M. 2004, \href{http://dx.doi.org/10.1126/science.1090720}{\JournalTitle{Science}, 304, 536}

\bibitem[{{Lorimer} {et~al.}(2006){Lorimer}, {Stairs}, {Freire}, {Cordes}, {Camilo}, {Faulkner}, {Lyne}, {Nice}, {Ransom}, {Arzoumanian}, {Manchester}, {Champion}, {van Leeuwen}, {Mclaughlin}, {Ramachandran}, {Hessels}, {Vlemmings}, {Deshpande}, {Bhat}, {Chatterjee}, {Han}, {Gaensler}, {Kasian}, {Deneva}, {Reid}, {Lazio}, {Kaspi}, {Crawford}, {Lommen}, {Backer}, {Kramer}, {Stappers}, {Hobbs}, {Possenti}, {D'Amico}, \& {Burgay}}]{Lorimer2006}
{Lorimer}, D.~R., {Stairs}, I.~H., {Freire}, P.~C., {et~al.} 2006, \href{http://dx.doi.org/10.1086/499918}{\JournalTitle{\apj}, 640, 428}

\bibitem[{{Lyne} \& {Graham-Smith}(2012)}]{Lyne2012}
{Lyne}, A. \& {Graham-Smith}, F. 2012, {Pulsar Astronomy}

\bibitem[{{Lyne} {et~al.}(2004){Lyne}, {Burgay}, {Kramer}, {Possenti}, {Manchester}, {Camilo}, {McLaughlin}, {Lorimer}, {D'Amico}, {Joshi}, {Reynolds}, \& {Freire}}]{Lyne2004}
{Lyne}, A.~G., {Burgay}, M., {Kramer}, M., {et~al.} 2004, \href{http://dx.doi.org/10.1126/science.1094645}{\JournalTitle{Science}, 303, 1153}

\bibitem[{{Maan} {et~al.}(2021){Maan}, {van Leeuwen}, \& {Vohl}}]{Maan2021}
{Maan}, Y., {van Leeuwen}, J., \& {Vohl}, D. 2021, \href{http://dx.doi.org/10.1051/0004-6361/202040164}{\JournalTitle{\aap}, 650, A80}

\bibitem[{{Manchester} {et~al.}(2005){Manchester}, {Hobbs}, {Teoh}, \& {Hobbs}}]{Manchester2005}
{Manchester}, R.~N., {Hobbs}, G.~B., {Teoh}, A., \& {Hobbs}, M. 2005, \href{http://dx.doi.org/10.1086/428488}{\JournalTitle{\aj}, 129, 1993}

\bibitem[{{Mikhailov} {et~al.}(2017){Mikhailov}, {van Leeuwen}, \& {Jonker}}]{Mikhailov2017}
{Mikhailov}, K., {van Leeuwen}, J., \& {Jonker}, P.~G. 2017, \href{http://dx.doi.org/10.3847/1538-4357/aa696a}{\JournalTitle{\apj}, 840, 9}

\bibitem[{{Ng} {et~al.}(2018){Ng}, {Kruckow}, {Tauris}, {Lyne}, {Freire}, {Ridolfi}, {Caiazzo}, {Heyl}, {Kramer}, {Cameron}, {Champion}, \& {Stappers}}]{Ng2018}
{Ng}, C., {Kruckow}, M.~U., {Tauris}, T.~M., {et~al.} 2018, \href{http://dx.doi.org/10.1093/mnras/sty482}{\JournalTitle{\mnras}, 476, 4315}

\bibitem[{{Noutsos} {et~al.}(2020){Noutsos}, {Desvignes}, {Kramer}, {Wex}, {Freire}, {Stairs}, {McLaughlin}, {Manchester}, {Possenti}, {Burgay}, {Lyne}, {Breton}, {Perera}, \& {Ferdman}}]{Noutsos2020}
{Noutsos}, A., {Desvignes}, G., {Kramer}, M., {et~al.} 2020, \href{http://dx.doi.org/10.1051/0004-6361/202038566}{\JournalTitle{\aap}, 643, A143}

\bibitem[{{{\"O}zel} \& {Freire}(2016)}]{Ozel2016}
{{\"O}zel}, F. \& {Freire}, P. 2016, \href{http://dx.doi.org/10.1146/annurev-astro-081915-023322}{\JournalTitle{\araa}, 54, 401}

\bibitem[{{Pastor-Marazuela} {et~al.}(2023){Pastor-Marazuela}, {Straal}, {van Leeuwen}, \& {Kondratiev}}]{Pastor2023}
{Pastor-Marazuela}, I., {Straal}, S.~M., {van Leeuwen}, J., \& {Kondratiev}, V.~I. 2023, \href{http://dx.doi.org/10.1051/0004-6361/202245214}{\JournalTitle{\aap}, 672, A151}

\bibitem[{{Perera} {et~al.}(2010){Perera}, {McLaughlin}, {Kramer}, {Stairs}, {Ferdman}, {Freire}, {Possenti}, {Breton}, {Manchester}, {Burgay}, {Lyne}, \& {Camilo}}]{Perera2010}
{Perera}, B.~B.~P., {McLaughlin}, M.~A., {Kramer}, M., {et~al.} 2010, \href{http://dx.doi.org/10.1088/0004-637X/721/2/1193}{\JournalTitle{\apj}, 721, 1193}

\bibitem[{{Pitkin}(2018)}]{psrqpy}
{Pitkin}, M. 2018, \href{http://dx.doi.org/10.21105/joss.00538}{\JournalTitle{{Journal of Open Source Software}}, 3, 538}

\bibitem[{{Ransom}(2011)}]{Ransom2011}
{Ransom}, S. 2011, {PRESTO: PulsaR Exploration and Search TOolkit}, Astrophysics Source Code Library, record ascl:1107.017

\bibitem[{{Ridolfi}(2020)}]{Ridolfi2020}
{Ridolfi}, A. 2020, {PYSOLATOR: Remove orbital modulation from a binary pulsar and/or its companion}, Astrophysics Source Code Library, record ascl:2003.012

\bibitem[{{Tauris} {et~al.}(2017){Tauris}, {Kramer}, {Freire}, {Wex}, {Janka}, {Langer}, {Podsiadlowski}, {Bozzo}, {Chaty}, {Kruckow}, {van den Heuvel}, {Antoniadis}, {Breton}, \& {Champion}}]{Tauris2017}
{Tauris}, T.~M., {Kramer}, M., {Freire}, P.~C.~C., {et~al.} 2017, \href{http://dx.doi.org/10.3847/1538-4357/aa7e89}{\JournalTitle{\apj}, 846, 170}

\bibitem[{{Taylor} {et~al.}(1976){Taylor}, {Hulse}, {Fowler}, {Gullahorn}, \& {Rankin}}]{Taylor76}
{Taylor}, J.~H., {Hulse}, R.~A., {Fowler}, L.~A., {Gullahorn}, G.~E., \& {Rankin}, J.~M. 1976, \href{http://dx.doi.org/10.1086/182131}{\JournalTitle{\apjl}, 206, L53}

\bibitem[{{van Kerkwijk} \& {Kulkarni}(1999)}]{vanKerkwijk1999}
{van Kerkwijk}, M.~H. \& {Kulkarni}, S.~R. 1999, \href{http://dx.doi.org/10.1086/311991}{\JournalTitle{\apjl}, 516, L25}

\bibitem[{{van Leeuwen} {et~al.}(2020){van Leeuwen}, {Mikhailov}, {Keane}, {Coenen}, {Connor}, {Kondratiev}, {Michilli}, \& {Sanidas}}]{vL2020}
{van Leeuwen}, J., {Mikhailov}, K., {Keane}, E., {et~al.} 2020, \href{http://dx.doi.org/10.1051/0004-6361/201937065}{\JournalTitle{\aap}, 634, A3}

\bibitem[{{van Leeuwen} \& {Stappers}(2010)}]{ls10}
{van Leeuwen}, J. \& {Stappers}, B.~W. 2010, \href{http://dx.doi.org/10.1051/0004-6361/200913121}{\JournalTitle{\aap}, 509, 7}

\bibitem[{{van Leeuwen} \& {Verbunt}(2004)}]{lv04}
{van Leeuwen}, J. \& {Verbunt}, F. 2004, in IAU Symposium, Vol. 218, Young Neutron Stars and Their Environments, ed. {F.~Camilo \& B.~M.~Gaensler}, 41

\bibitem[{{van Leeuwen} {et~al.}(2006){van Leeuwen}, {Cordes}, {Lorimer}, {Freire}, {Camilo}, {Stairs}, {Nice}, {Champion}, {Ramachandran}, {Faulkner}, {Lyne}, {Ransom}, {Arzoumanian}, {Manchester}, {McLaughlin}, {Hessels}, {Vlemmings}, {Deshpande}, {Bhat}, {Chatterjee}, {Han}, {Gaensler}, {Kasian}, {Deneva}, {Reid}, {Lazio}, {Kaspi}, {Crawford}, {Lommen}, {Backer}, {Kramer}, {Stappers}, {Hobbs}, {Possenti}, {D'Amico}, {Faucher-Gigu{\`e}re}, \& {Burgay}}]{lcl+06}
{van Leeuwen}, J., {Cordes}, J.~M., {Lorimer}, D.~R., {et~al.} 2006, \href{http://dx.doi.org/10.1088/1009-9271/6/S2/58}{\JournalTitle{Chin. J. Astron. Astrophys.}, 6, 020}

\bibitem[{{van Leeuwen} {et~al.}(2015){van Leeuwen}, {Kasian}, {Stairs}, {Lorimer}, {Camilo}, {Chatterjee}, {Cognard}, {Desvignes}, {Freire}, {Janssen}, {Kramer}, {Lyne}, {Nice}, {Ransom}, {Stappers}, \& {Weisberg}}]{vanLeeuwen2015}
{van Leeuwen}, J., {Kasian}, L., {Stairs}, I.~H., {et~al.} 2015, \href{http://dx.doi.org/10.1088/0004-637X/798/2/118}{\JournalTitle{\apj}, 798, 118}

\bibitem[{{Venkatraman Krishnan} {et~al.}(2019){Venkatraman Krishnan}, {Bailes}, {van Straten}, {Keane}, {Kramer}, {Bhat}, {Flynn}, \& {Os{\l}owski}}]{VenkatramanKrishnan2019}
{Venkatraman Krishnan}, V., {Bailes}, M., {van Straten}, W., {et~al.} 2019, \href{http://dx.doi.org/10.3847/2041-8213/ab0a03}{\JournalTitle{\apjl}, 873, L15}

\bibitem[{{Venkatraman Krishnan} {et~al.}(2020){Venkatraman Krishnan}, {Bailes}, {van Straten}, {Wex}, {Freire}, {Keane}, {Tauris}, {Rosado}, {Bhat}, {Flynn}, {Jameson}, \& {Os{\l}owski}}]{VenkatramanKrishnan2020}
{Venkatraman Krishnan}, V., {Bailes}, M., {van Straten}, W., {et~al.} 2020, \href{http://dx.doi.org/10.1126/science.aax7007}{\JournalTitle{Science}, 367, 577}

\bibitem[{{Vleeschower}(2024)}]{Vleeschower2024}
{Vleeschower}, L. 2024, PhD thesis, The University of Manchester, {\url{https://research.manchester.ac.uk/en/studentTheses/searching-for-pulsars-for-testing-gravity-with-the-ska-precursor-}}

\bibitem[{{Wang}(2025)}]{WangThesis2025}
{Wang}, Y.~Y. 2025, PhD thesis, The University of Amsterdam

\bibitem[{{Willems} \& {Kalogera}(2004)}]{Willems2004}
{Willems}, B. \& {Kalogera}, V. 2004, \href{http://dx.doi.org/10.1086/383200}{\JournalTitle{\apjl}, 603, L101}

\bibitem[{{Yang} {et~al.}(2017){Yang}, {Zhang}, {Li}, {Wang}, {Pan}, {Lingfu}, \& {Zhou}}]{YangYY2017}
{Yang}, Y.-Y., {Zhang}, C.-M., {Li}, D., {et~al.} 2017, \href{http://dx.doi.org/10.3847/1538-4357/835/2/185}{\JournalTitle{\apj}, 835, 185}

\end{thebibliography}

\end{document}